\documentclass[12pt]{article}
\usepackage{graphics}
\begin{document}
\begin{flushright}
WISC-MILW-98-TH-17 \\
{\em European Journal of Physics, vol. 19, pg. 143-150 (1998)}\\
\end{flushright}
\vspace{.4in}
\begin{center}
{\LARGE From Newton's Laws to the}\\
\vspace{.25in}
{\LARGE Wheeler-DeWitt Equation}\\
\vspace{.6in}
{\Large John W. Norbury}\\
\vspace{.1in}
{\em Physics Department, University of Wisconsin - Milwaukee, \\
P.O. Box 413, Milwaukee, Wisconsin 53201,
USA}\\ (e-mail: norbury@uwm.edu)\\
\vspace{1in}

{\bf Abstract}
\end{center}
This is a pedagogical paper which explains some ideas in cosmology at a level accessible to
undergraduate students. It does not use general relativity, but uses the ideas of Newtonian
cosmology worked out by Milne and McCrea. The cosmological constant is
also introduced within a Newtonian framework. Following standard quantization procedures 
the Wheeler-DeWitt equation in the minisuperspace approximation is derived for empty and non-empty universes.
\vspace{5mm}

\newpage
\section{Introduction}

Some of the modern ideas in cosmology can be explained without the need to discuss general
 relativity [Landsberg and Evans (1977)]. The present paper represents an attempt to do this based on
Newtonian mechanics.  There is a need for pedagogical articles which discuss forefront ideas in research, but
also include all the relevant derivations {\em in one place} and at an accessible level. This will
 be useful for instructors not familiar with cosmology but who would nevertheless like a 
summary  in a single article that can be used in, say, a modern physics
course.  There are also some ideas presented here that cannot be found in the pedagogical
literature. These include a discussion of how to incorporate the cosmological constant in Newtonian
mechanics and a discussion of the Wheeler-DeWitt equation for flat, open and closed empty and non-empty
universes. 

\section{Equation of State}
In what follows the equation of state for matter and radiation will be needed. 
In particular an expression for the rate of change of density, $\dot {\rho}$, will be needed in terms of
the density $\rho $ and pressure $p$. (The definition $\dot {x} \equiv \frac{dx}{dt}$, where $t$ is time,
is being used.) The first law of thermodynamics is
\begin{equation}
d{\bar U} + dW = dQ
\end{equation}
where ${\bar U}$ is the internal energy, $W$ is the work and $Q$ is the heat transfer. Ignoring any 
heat transfer and writing $dW = F dr = p dV$ where $F$ is the force, $r$ is the distance, $p$ is the
pressure and $V$ is the volume, then
\begin{equation}
d{\bar U} = -p dV .          \label{eq:state 1}
\end{equation}
Assuming that $\rho$ is a relativistic energy density means that the energy is expressed as [Guth and
Steinhardt (1989)] 
\begin{equation}
{\bar U} = \rho V
\end{equation}
from which it follows that
\begin{equation}
\dot{\bar U} = \dot{\rho} V + \rho \dot{V} = -p \dot{V}
\end{equation}
where the term on the far right hand side results from equation~(\ref{eq:state 1}). 
Writing $V \propto r^3$ implies that $\frac{\dot{V}}{V} = 3 \frac{\dot{r}}{r} $. Thus
\begin{equation}
\dot{\rho } = -3(\rho + p) \frac{\dot{r}}{r}      \label{eq:rhodot}
\end{equation}

\subsection{Matter}
Writing the density of matter as  
\begin{equation}
\rho = \frac {M}{\frac{4}{3}\pi r^3}        \label{eq:dense}
\end{equation}
it follows that
\begin{equation}
\dot{\rho} \equiv \frac{d \rho}{dr} \dot{r} = -3\rho \frac{\dot{r}}{r}
\end{equation}
so that by comparing to equation~(\ref{eq:rhodot}), it follows that the equation of state for 
matter is
\begin{equation}
p=0.
\end{equation}
This is the same as obtained from the ideal gas law for zero temperature. Recall that in this
 derivation we have not introduced any kinetic energy, so we are talking about zero temperature.

\subsection{Radiation}
The equation of state for radiation  can be derived by considering radiation modes in a cavity  
based on analogy with a violin string [Kubo (1967)]. For a standing wave on a string fixed at both
ends
\begin{equation}
L=\frac{n \lambda}{2} 
\end{equation}
where $L$ is the length of the string,  $\lambda$ is the wavelength and $n$ is a positive integer 
($n=1,2,3 .....$). Radiation travels at the velocity of light, so that
\begin{equation}
c=f \lambda = f \frac{2L}{n}
\end{equation}
where $f$ is the frequency. Thus substituting $f = \frac{n}{2L} c$ into Planck's 
formula ${\bar U}= \hbar \omega = h f$, where $h$ is Planck's constant, gives
\begin{equation}
{\bar U}=\frac{n h c}{2} \frac{1}{L} \propto V^{-1/3}.
\end{equation}
Using equation~(\ref {eq:state 1}) the pressure becomes
\begin{equation}
p \equiv - \frac{d{\bar U}}{dV} = \frac{1}{3} \frac{{\bar U}}{V} .
\end{equation}
Using $\rho = {\bar U} / V $, the radiation equation of state is 
\begin{equation}
p = \frac{1}{3} \rho .
\end{equation}
It is customary to combine the equations of state into the form
\begin{equation}
p = \frac{\gamma}{3} \rho       \label{eq:eqn state}
\end{equation}
where $\gamma \equiv 1$ for radiation and $\gamma \equiv 0$ for matter. These equations of
state are needed in order to discuss the radiation and matter dominated epochs which occur in the
evolution of the Universe.

\section{Velocity and Acceleration Equations}

The  equation which specifies the speed of recession is obtained by
writing the  total energy $E$ as the sum of kinetic plus potential energy terms (and using
$M=\frac{4}{3}\pi r^3\rho$) [Madsen (1995), Roos (1994)] 
\begin{equation}
E=T+{\tilde U}= \frac{1}{2} m \dot{r}^2 - G \frac{M m}{r} = \frac{1}{2} m r^2 (H^2 - 
\frac{8 \pi G}{3} \rho)     \label{eq:energy}    
\end{equation}
where the Hubble constant $H \equiv \frac{\dot{r}}{r}$, $m$ is the mass of a 
test particle in the potential energy field enclosed by a gas of dust of mass $M$, $r$ is the distance
from the center of the dust to the test particle and $G$ is Newton's constant. Recall that the escape
velocity is just $v_{escape}=\sqrt{\frac{2GM}{r}}=\sqrt{\frac{8\pi G}{3} \rho r^2}$, so that the above
equation can also be written
\begin{equation}
\dot{r}^2 =  v_{escape}^2 - k^{\prime}         
\end{equation}
with $k^{\prime} \equiv -\frac{2E}{m}$. 
The constant $k^{\prime}$ can either be negative, zero or positive corresponding 
to the total energy $E$ being positive, zero or negative. For a particle in motion near the Earth this
would correspond to the particle escaping (unbound), reaching infinity with zero speed (critical case) or
returning (bound) to Earth because the speed $\dot{r}$ would be greater, equal to or smaller than the
escape speed
$v_{escape}$. Later this will be analagous to an open, flat or closed universe.
Equation~(\ref{eq:energy}) is re-arranged as
\begin{equation}
H^2=\frac{8\pi G}{3} \rho +  \frac{2E}{m r^2} .
\end{equation}
Defining $k \equiv -\frac{2E}{m s^2}$ and writing the distance in terms of a scale 
factor $a$ and a constant length $s$ as $r(t) \equiv a(t) s$, it follows that $\frac{\dot{r}}{r} =
\frac{\dot{a}}{a}$ and $\frac{\ddot{r}}{r} = \frac{\ddot{a}}{a}$, giving  [Madsen
(1995), Roos (1994)]
\begin{equation}
H^2 \equiv (\frac{\dot{a}}{a})^2=\frac{8\pi G}{3}\rho-\frac{k}{a^2}         \label{eq:F00a}
\end{equation}
which specifies the speed of recession. The scale factor is introduced because in 
general relativity it is space itself which expands. Even though this equation is
derived for matter, it is also true for radiation.   The same equation is
obtained in general relativity [Islam (1992)].  According to Guth
[Guth and Steinhardt (1989)], k can be rescaled so that instead of being negative, zero or positive
it takes on the values $-1, 0$ or $+1$.  From a geometric, general relativistic point of view
this corresponds to an open, flat or closed universe. 

In elementary mechanics the speed $v$ of a ball dropped from a height $r$ is evaluated from 
the conservation of energy equation as $v=\sqrt{2gr}$, where $g$ is the acceleration due to gravity. The
derivation shown above is exactly analagous to such a calculation. Similarly the acceleration $a$ of the
ball is calculated as $a=g$ from Newton's equation $F=m\ddot{r}$, where $F$ is the force and the
acceleration is $\ddot{r} \equiv \frac{d^{2}r}{dt^2}$. The acceleration for the universe is obtained from
Newton's equation
\begin{equation}
-G\frac{Mm}{r^2} = m \ddot{r} .  
\end{equation}
Again using  $M=\frac{4}{3} \pi r^3 \rho$ and $\frac{\ddot{r}}{r} = \frac{\ddot{a}}{a}$ gives 
the acceleration equation
\begin{equation}
\frac{F}{m r} \equiv \frac{\ddot{r}}{r} \equiv
\frac{\ddot{a}}{a}=-\frac{4\pi G}{3}\rho .   \label{eq:Fii1}    
\end{equation}
However because $M=\frac{4}{3} \pi r^3 \rho$ was used, it is clear that this acceleration 
equation holds only for matter. In our example of the falling ball instead of the acceleration being
obtained from Newton's Law, it can also be obtained by taking the time derivative of the energy equation
to give $a=\frac{dv}{dt}= v \frac{dv}{dr} = (\sqrt{2gr}) ( \sqrt{2g} \frac{1}{2 \sqrt{r}}) = g$.
Similarly, for the general case one can take the time derivative of equation~(\ref{eq:F00a}) (valid for
matter and radiation) [Madsen (1995)]
\begin{equation}
\frac{d}{dt} \dot{a}^2  = 2\dot{a} \ddot{a} = \frac{8\pi G}{3}\frac{d}{dt} (\rho a^2) .    
\end{equation}
Upon using equation~(\ref{eq:rhodot}) the acceleration equation is obtained as
\begin{equation}
\frac{\ddot{a}}{a}=-\frac{4\pi G}{3}(\rho + 3p)
=-\frac{4\pi G}{3}(1+\gamma)\rho              \label{eq:Fiia}
\end{equation}
which reduces to equation~(\ref{eq:Fii1}) for the matter equation of state ($\gamma = 0$). 
The same equation is obtained in general relativity [Islam (1992), Milne (1934),
McCrea and Milne (1934), Bondi (1961)].

\section{Cosmological Constant}

In both Newtonian and relativistic cosmology the universe is unstable to gravitational collapse. 
Both Newton and Einstein believed that the Universe is static. In order to obtain this Einstein
introduced a {\it repulsive} gravitational force, called the cosmological constant, and Newton could have
done exactly the same thing, had he believed the universe to be finite.

In order to obtain a possibly zero acceleration, a positive term (conventionally taken 
as $\frac{\Lambda }{3}$ ) is added to the acceleration equation~(\ref{eq:Fiia}) as
\begin{equation}
\frac{\ddot{a}}{a}=-\frac{4\pi G}{3}(\rho + 3p)
+\frac{\Lambda }{3}              \label{eq:Fiicc}
\end{equation}
which, with the proper choice of $\Lambda $ can give the required zero acceleration for a 
static universe. Again exactly the same equation is obtained from the Einstein field equations
[Islam (1992)]. What has been done here is entirely equivalent to just adding a repulsive
gravitational force in Newton's Law. The question now is how this repulsive force enters the energy
equation~(\ref{eq:F00a}). Identifying the force from 
\begin{equation}
\frac{\ddot{r}}{r} = \frac{\ddot{a}}{a} \equiv 
\frac{F_{repulsive}}{m r} \equiv \frac{\Lambda }{3}
\end{equation}
and using 
\begin{equation}
F_{repulsive} =  \frac{\Lambda }{3} m r \equiv - \frac{d{\tilde U}}{dr}  
\end{equation}
gives the potential energy as 
\begin{equation}
{\tilde U}_{repulsive} =  - \frac{1}{2} \frac{\Lambda }{3} m r^2  
\end{equation}
which is just a {\it repulsive} simple harmonic oscillator. Substituting this into the
 conservation of energy equation
\begin{equation}
E=T+{\tilde U}=\frac{1}{2} m \dot{r}^2 - G \frac{M m}{r} - \frac{1}{2} \frac{\Lambda }{3} m r^2 
= \frac{1}{2} m r^2(H^2-\frac{8\pi G}{3} \rho - \frac{\Lambda }{3})        
\end{equation}
gives
\begin{equation}
H^2 \equiv (\frac{\dot{a}}{a})^2=\frac{8\pi G}{3}\rho-\frac{k}{a^2} + \frac{\Lambda }{3}  .     
  \label{eq:F00cc}
\end{equation}

Let us comment on the repulsive harmonic oscillator obtained above. Recall one of the standard 
problems often assigned to students in mechanics courses. The problem is to imagine that a hole has been
drilled from one side of the Earth, through the center and to the other side and to show that if a
ball is dropped into the hole, it will execute harmonic motion. The solution is obtained by noting that
whereas gravity is an inverse square law for point masses $M$ and $m$ separated by a distance $r$ as
given by $F=G\frac{M m}{r^2}$, yet if one of the masses is a continous mass distribution represented by a
density then $F=G\frac{4}{3}\pi \rho m r$. The force rises linearly as the distance is increased because
the amount of matter enclosed keeps increasing. Thus the gravitational force for a continuous mass
distribution rises like Hooke's law and thus oscillatory solutions are encountered. This sheds light on
our repulsive oscillator found above. In this case we want the gravity to be repulsive, but the
cosmological constant acts just like the uniform matter distribution. 

Finally authors often write the cosmological constant in terms of a vacuum energy density 
as $\Lambda \equiv 8\pi G \rho_{vac}$ so that the velocity and acceleration equations become

\begin{equation}
H^2 \equiv (\frac{\dot{a}}{a})^2=\frac{8\pi G}{3}\rho-\frac{k}{a^2}+\frac{\Lambda}{3}
=\frac{8\pi G}{3}(\rho + \rho_{vac})-\frac{k}{a^2}
\label{eq:F00}
\end{equation}
and
\begin{equation}
\frac{\ddot{a}}{a}=-\frac{4\pi G}{3}(1+\gamma)\rho +\frac{\Lambda}{3}
=-\frac{4\pi G}{3}(1+\gamma)\rho +\frac{8\pi G}{3}\rho_{vac} .
\label{eq:Fii}
\end{equation}

\subsection{Einstein Static Universe}

Although we have noted that the cosmological constant provides repulsion, it is interesting to 
calculate its exact value for a static universe [Atwater (1974), Adler et al (1975)]. The Einstein
static universe requires $a=a_{0}=constant$ and thus $\dot{a}=\ddot{a}=0$. The case $\ddot{a}=0$ will be
examined first. From equation~(\ref{eq:Fiicc}) this requires that
\begin{equation}
\Lambda = 4\pi G (\rho + 3p) = 4\pi G (1+\gamma )\rho .
\label{eq:static 1}
\end{equation}
If there is no  cosmological constant ($\Lambda=0$) then either $\rho = 0$ which is an empty 
universe, or $p=-\frac{1}{3} \rho$ which requires negative pressure. Both of these alternatives were
unacceptable to Einstein and therefore he concluded that a cosmological constant was present, i.e.
$\Lambda \neq 0$. From equation~(\ref{eq:static 1}) this implies
\begin{equation}
\rho = \frac{\Lambda}{4\pi G (1+\gamma )}
\label{eq:static 2}
\end{equation}
and because $\rho$ is positive this requires a positive $\Lambda$. Substituting 
equation~(\ref{eq:static 2}) into equation~(\ref{eq:F00cc}) it follows that
\begin{equation}
\Lambda = \frac{3(1+\gamma )}{3+\gamma}[(\frac{\dot{a}}{a_{0}})^2 + \frac{k}{a_{0}^2}] .
\label{eq:int}
\end{equation}
Now imposing $\dot{a}=0$ and assuming a matter equation of state ($\gamma =0$) 
implies $\Lambda = \frac{k}{a_{0}^2}$. However the requirement that $\Lambda$ be positive forces $k=+1$
giving
\begin{equation}
\Lambda=\frac{1}{a_{0}^2}=constant .
\label{eq:Einstein}
\end{equation}
Thus the  cosmological constant is not any old value but rather simply the inverse of the scale
 factor squared, where the scale factor has a fixed value in this static model.

\section{Conservation laws}

Just as the Maxwell equations imply the conservation of charge, so too do our velocity and
 acceleration equations imply conservation of energy. The energy-momentum conservation equation is
derived by setting the covariant derivative of the energy momentum tensor equal to zero. The same result
is achieved by taking the time derivative of equation~(\ref{eq:F00}). The result is
\begin{equation}
\dot{\rho} +3 (\rho + p) \frac{\dot{a}}{a}=0 .
\label{eq:cons 1}
\end{equation}
or
\begin{equation}
\frac{d}{dt} (\rho a^3) + p \frac{da^3}{dt}=0
\label{eq:cons 2}
\end{equation}
and from equation~(\ref{eq:eqn state}), $3(\rho + p)=(3+\gamma )\rho$, it follows that
\begin{equation}
\frac{d}{dt} (\rho a^{3+\gamma})  =0.
\label{eq:cons 3}
\end{equation}
Integrating this we obtain
\begin{equation}
\rho=\frac{A}{a^{3+\gamma}}
\label{eq:rate soln}
\end{equation}
where $A$ is a constant given by $A \equiv \rho_0 a_0^{3+\gamma}$. This shows that the density falls as
$\frac{1}{a^3}$ for matter  and $\frac{1}{a^4}$ for radiation as expected.

Later we shall use these equations in a different form as follows.
From equation~(\ref{eq:cons 1}),
\begin{equation}
\rho^\prime +3 (\rho + p) \frac{1}{a} =0
\label{eq:cons 4}
\end{equation}
where primes denote derivatives with respect to a, i.e. $x^\prime \equiv dx/da$. Alternatively
\begin{equation}
\frac{d}{da} (\rho a^3) + 3pa^2  =0
\label{eq:cons 5}
\end{equation}
so that 
\begin{equation}
\frac{1}{a^{3+\gamma}} \frac{d}{da} (\rho a^{3+\gamma}) = 0
\label{eq:rate}
\end{equation}
which is consistent with equation~(\ref{eq:rate soln})

\section{Quantum Cosmology}
The discussion of the Wheeler-DeWitt equation in the minisuperspace approximation 
[Hartle and Hawking (1983), Kolb and Turner (1990), Atkatz (1994), Atkatz and Pagels (1982)] is
usually restricted to closed ($k=+1$) and empty ($\rho = 0$) universes. Atkatz [Atkatz (1994)]
presented a very nice discussion for closed and empty universes. Herein we consider closed, open and flat
and empty and non-empty universes. It is important to consider the possible presence of matter and
radiation as they might otherwise change the conclusions. Thus presented below is a derivation of the
Wheeler-DeWitt equation in the minisuperspace approximation which also includes matter and radiation and
arbitrary values of $k$.

The Lagrangian is
\begin{equation}
L=-\frac{\kappa}{k^{3/2}} a^3 [ (\frac{\dot{a}}{a})^2 -\frac{k}{a^2} + \frac{8\pi G}{3}(\rho + \rho_{vac})
]
\label{eq:L}
\end{equation}
with $\kappa \equiv \frac{3\pi }{4G}$. Note that this Lagrangian is not defined for a flat ($k=0$)
universe. Also for an open universe ($k=-1$) the Lagrangian is complex ($k^{3/2} = -i$).

The momentum conjugate to a is 
\begin{equation}
P \equiv \frac{\partial L}{\partial \dot{a}}=-\frac{\kappa}{k^{3/2}} 2 a \dot{a} .
\label{eq:momn}
\end{equation}
Substituting L and P into the Euler-Lagrange equation, $\dot{P}-\frac{\partial L}{\partial a}=0$, 
equation~(\ref{eq:F00}) is recovered. (Note the calculation of $\frac{\partial L}{\partial a}$ is
simplified by using the conservation equation~(\ref{eq:cons 4}) with equation~(\ref{eq:eqn state}),
namely $\rho^\prime   = - (3+\gamma) \rho / a$). The Hamiltonian ${\bar H} \equiv
P\dot{a} - L$ is
\begin{equation}
{\bar H}(\dot{a}, a)=-\frac{\kappa}{k^{3/2}} a^3 [ (\frac{\dot{a}}{a})^2 +\frac{k}{a^2} - \frac{8\pi
G}{3}(\rho + 
\rho_{vac}) ] \equiv  0
\label{eq:Ha}
\end{equation}
which has been written in terms of $\dot{a}$ to show explicitly that the Hamiltonian is 
identically zero and is not equal to the total energy as before. (Compare equation~(\ref{eq:F00})). In
terms of the conjugate momentum
\begin{equation}
{\bar H}(P, a)=-\frac{\kappa}{k^{3/2}} a^3 [ \frac{k^3}{4\kappa ^2 a^4}P^2 +\frac{k}{a^2} - \frac{8\pi
G}{3}(\rho +
 \rho_{vac}) ] = 0
\label{eq:HP}
\end{equation}
which, of course is also equal to zero. Making the replacement $P \rightarrow -i\frac{\partial }
{\partial a}$ and imposing ${\bar H}\Psi =0$ results in the Wheeler-DeWitt equation in the minisuperspace
approximation  for arbitrary $k$ and with matter or radiation ($\rho$ term) included gives
\begin{equation}
\{-\frac{d^2}{d a^2}+\frac{9\pi ^2}{4G^2 k^2}[(a^2-\frac{8\pi G}{3 k}(\rho + 
\rho_{vac})a^4]\}\Psi=0  .
\label{eq:wdwe}
\end{equation}
Using equation~(\ref{eq:rate soln}) the Wheeler-DeWitt equation becomes
\begin{equation}
\{-\frac{d^2}{d a^2}+\frac{9\pi ^2}{4G^2 k^2}[a^2-\frac{1}{k}(\frac{\Lambda}{3}a^4+
\frac{8\pi G}{3}Aa^{1-\gamma})]\}\Psi=0  .
\label{eq:wdwe1}
\end{equation}
This just looks like the zero energy Schr\"{o}dinger equation [Kolb and Turner (1990)] with a
potential given by 
\begin{equation}
U(a)=\frac{9\pi ^2}{4G^2 k^2}[ a^2-\frac{1}{k}(\frac{\Lambda}{3}a^4+
\frac{8\pi G}{3}Aa^{1-\gamma})] .
\label{eq:pot1}
\end{equation}

 For the empty Universe case of no matter or radiation ($A=0$) the
potential
$U(a)$ is plotted  in Figure 1 for the cases $k=+1,  -1$ respectively corresponding to closed [Kolb and
Turner (1990)] and open universes. It can be seen that only the closed universe case provides a
potential barrier through which tunneling can occur. The open universe would immediately recollapse.
(Actually for the open universe the Lagrangian is complex, so this discussion may be meaningless. In
either case it is clear that an open universe cannot arise via quantum tunneling.) This provides a clear
illustration of the idea that only closed universes can arise through quantum tunneling [Atkatz and
Pagels (1982)]. For a flat universe ($k=0$), the potential is not defined and thus this process of
quantum tunneling cannot give rise to a flat universe. If radiation ($\gamma = 1$ and $A
\neq 0$) is included then only a negative constant will be added to the potential (because the term
$a^{1-\gamma}$ will be constant for
$\gamma = 1$) and these conclusions about tunneling will not change. The shapes in Figure 1 will be
identical except that the whole graph will be shifted downwards by a constant with the inclusion of
radiation. (For matter ($\gamma = 0$ and $A \neq 0$) a term growing like $a$ will be included in the
potential which will only be important for very small $a$ and so the conclusions again will not be
changed.) To summarize, only closed universes can arise from quantum tunneling even if  matter or
radiation is present.

This work was supported  by the Wisconsin Space Grant Consortium.

\newpage

\includegraphics{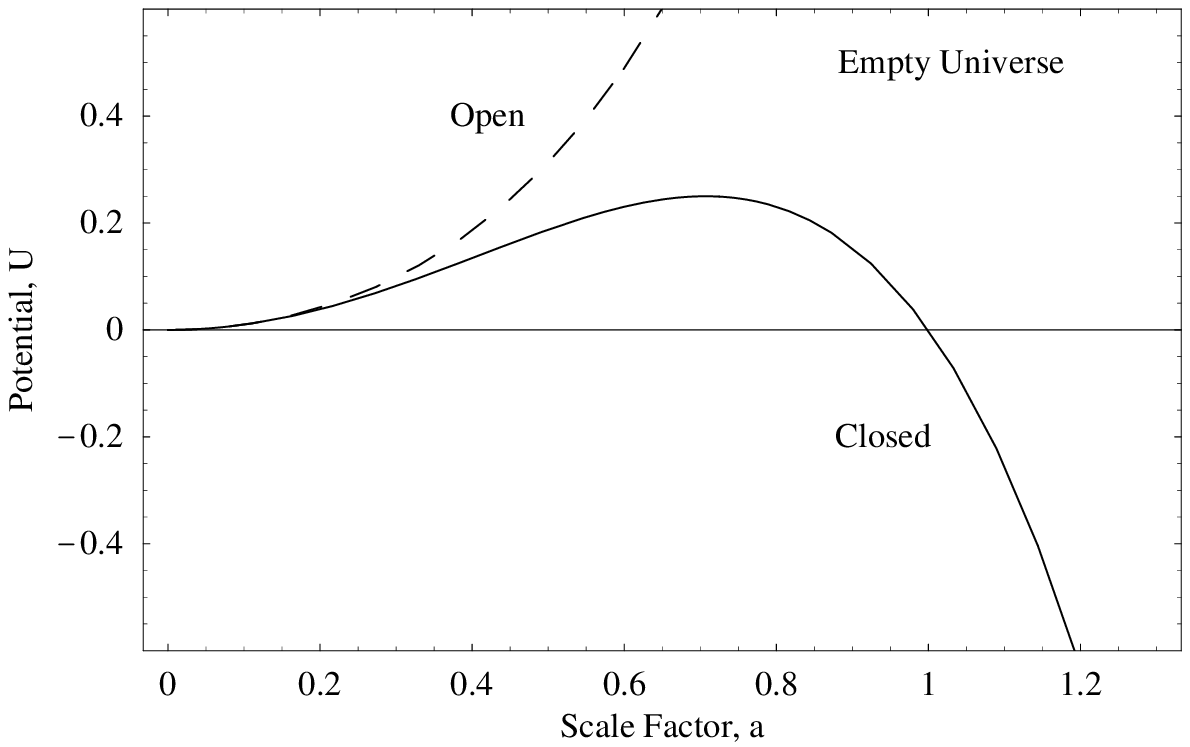}

\vspace{10mm} \noindent
\hspace{50mm}{\bf Figure  1}\\

\vspace{3mm} \noindent 
\hspace{5mm} Wheeler-DeWitt potential for open and closed universe.

\end{document}